\PassOptionsToPackage{unicode}{hyperref}
\PassOptionsToPackage{hyphens}{url}
\documentclass[
]{article}
\usepackage[a4paper, total={6in, 8in}]{geometry}
\usepackage{wrapfig}
\usepackage{blindtext}
\usepackage{multicol}
\usepackage{lmodern}
\usepackage{authblk}
\usepackage{amssymb,amsmath}
\usepackage{ifxetex,ifluatex}
\ifnum 0\ifxetex 1\fi\ifluatex 1\fi=0 
  \usepackage[T1]{fontenc}
  \usepackage[utf8]{inputenc}
  \usepackage{textcomp} 
\else 
  \usepackage{unicode-math}
  \defaultfontfeatures{Scale=MatchLowercase}
  \defaultfontfeatures[\rmfamily]{Ligatures=TeX,Scale=1}
\fi
\IfFileExists{upquote.sty}{\usepackage{upquote}}{}
\IfFileExists{microtype.sty}{
  \usepackage[]{microtype}
  \UseMicrotypeSet[protrusion]{basicmath} 
}{}
\makeatletter
\@ifundefined{KOMAClassName}{
  \IfFileExists{parskip.sty}{%
    \usepackage{parskip}
  }{
    \setlength{\parindent}{0pt}
    \setlength{\parskip}{6pt plus 2pt minus 1pt}}
}{
  \KOMAoptions{parskip=half}}
\makeatother
\usepackage{xcolor}
\IfFileExists{xurl.sty}{\usepackage{xurl}}{} 
\IfFileExists{bookmark.sty}{\usepackage{bookmark}}{\usepackage{hyperref}}
\hypersetup{
  pdftitle={Structured Like a Language Model: Analysing AI as an Automated Subject},
  hidelinks,
  pdfcreator={LaTeX via pandoc}}
\usepackage{graphicx}
\makeatletter
\def\maxwidth{\ifdim\Gin@nat@width>\linewidth\linewidth\else\Gin@nat@width\fi}
\def\maxheight{\ifdim\Gin@nat@height>\textheight\textheight\else\Gin@nat@height\fi}
\makeatother
\setkeys{Gin}{width=\maxwidth,height=\maxheight,keepaspectratio}
\makeatletter
\def\fps@figure{htbp}
\makeatother
\title{Structured Like a Language Model: Analysing AI as an Automated Subject}

\author[1]{Liam Magee}
\author[2]{Vanicka Arora}
\author[3]{Luke Munn}
\affil[1]{Western Sydney University, Australia \authorcr l.magee@westernsydney.edu.au}
\affil[2]{University of Stirling, United Kingdom \authorcr vanicka.arora@stir.ac.uk}
\affil[3]{University of Queensland, Australia \authorcr l.munn@uq.edu.au}

\date{December 2022}

\begin{document}

\maketitle

\begin{abstract}
	Drawing from the resources of psychoanalysis and critical media
studies, in this paper we develop an analysis of Large Language Models
(LLMs) as `automated subjects'. We argue the intentional fictional
projection of subjectivity onto LLMs can yield an alternate frame
through which AI behaviour, including its productions of bias and harm,
can be analysed. First, we introduce language models, discuss their
significance and risks, and outline our case for interpreting model
design and outputs with support from psychoanalytic concepts. We trace a
brief history of language models, culminating with the releases, in
2022, of systems that realise `state-of-the-art' natural language
processing performance. We engage with one such system, OpenAI's
InstructGPT, as a case study, detailing the layers of its construction
and conducting exploratory and semi-structured interviews with chatbots.
These interviews probe the model's moral imperatives to be `helpful',
`truthful' and `harmless' by design. The model acts, we argue, as the
condensation of often competing social desires, articulated through the
internet and harvested into training data, which must then be regulated
and \emph{repressed}. This foundational structure can however be
redirected via prompting, so that the model comes to \emph{identify
with}, and \emph{transfer}, its commitments to the immediate human
subject before it. In turn, these automated productions of language can
lead to the human subject \emph{projecting} agency upon the model,
effecting occasionally further forms of \emph{countertransference}. We
conclude that critical media methods and psychoanalytic theory together
offer a productive frame for grasping the powerful new capacities of
AI-driven language systems.

\end{abstract}

\pagebreak

  \emph{Once the structure of language has been recognized in the
  unconscious, what sort of subject can we conceive for it? (Lacan, 2007)}

\begin{multicols*}{2}
  
  \hypertarget{introduction}{%
  \subsection{Introduction}\label{introduction}}
  
  In the few years since OpenAI's release of GPT-3, large language models
  (LLMs) -- neural networks that calculate probable textual sequences to
  follow linguistic input -- have become part of the infrastructural
  fabric for language-intensive software services in communications,
  advertising, healthcare, and IT. Refinements of GPT-3 (Codex, GPT-3.5,
  InstructGPT, ChatGPT) have stretched its capabilities for code
  generation, summarisation, translation, and question answering. Among
  social media subcommunities, LLMs have inspired degrees of creative
  experimentation, philosophical debate, and social critique unusual to
  nascent technologies. While scepticism toward the promise of LLM-based
  AI is common to these channels, critical scholarship supplies a more
  forensic insight into their limits. Bender et al. (2021) for example
  have described these models as `stochastic parrots': automatons able to
  stitch together probabilistic word continuations to form seemingly
  coherent and legible texts that are nonetheless devoid of context,
  intent or understanding. Kapoor and Narayanan (2022) have argued that
  journalistic accounts have downplayed model limitations, while Zhang
  \emph{et al.} (2022) suggest the vaunted increases in model scale can
  inhibit an ability to generalise and transfer knowledge to novel
  domains.
  
  These and other sanguine accounts (e.g., Leufer, 2020) of a technology
  too routinely hyped as verging upon sentience (Lemoine, 2022),
  artificial general intelligence (Fei et al., 2022) or the replacement of
  human labour for many language-based tasks, instead centre upon
  deficiencies that are presently, or perhaps inherently, embedded in
  these models. Numerous studies highlight problems of bias across gender,
  race, class, disability, and other categories (Abid et al., 2021; Bender
  et al., 2021; Bolukbasi et al., 2016; Buolamwini and Gebru, 2018; Hovy
  and Spruit, 2016; Kim et al., 2020; Magee et al., 2021; Whittaker et
  al., 2019), which translate into social harms and inequalities as they
  are rushed into production. A welcome effect of such studies is the
  regularity with which model authors now themselves include tests,
  analyses and mitigation strategies for correcting bias and minimising
  harm in the technical papers that accompany new releases. However, these
  remain framed predominantly within the epistemological horizons of
  technical disciplines. As critics note, applied in isolation,
  metrics-based evaluations (Liang et al., 2022) reinforce rather than
  remediate the structural conditions under which technologies like LLMs
  are developed and deployed. In more direct terms, questions about the
  demands for human labour, choice of textual sources, methods of
  operationalisation and end-uses of LLMs are rarely addressed in the
  technical literature. The fetish of technical performativity obscures
  the background engineering, commercial imperatives, and social
  orchestrations required to make these parrots talk.
  
  At least provisionally though, we depart from Bender \emph{et al.}'s
  (2021) account with respect to its implied, and necessarily reductive,
  ontological demarcation between human and machine. Our reasons are
  twofold. First, we follow emerging interest in language models as
  \emph{methodological} aids and provocateurs for qualitative research
  (Munk et al., 2022; Rettberg, 2022). Even if language models can be
  described as parrots with randomness, that does not seem to inhibit, at
  certain scales, a remarkable versatility and expressivity, which
  presents a potential for social and humanities research to consider its
  methods of investigating and interpreting these technical artefacts.
  Second, we conjecture that a counterfactual and intentional projection
  of subjectivity~onto LLMs -- not, as we qualify, more vaguely humanistic
  properties of sentience or consciousness -- can help to articulate other
  avenues for addressing bias and harm. Our interest here shifts from
  essential questions of identification and mitigation -- as LLMs
  themselves become attuned to these questions -- to those posed to the
  discursive presentation of an automated subjectivity: what exhibits
  itself through tone, style, deflection and so on as a structure that
  might be queried and contested.
  
  As an alternative to understanding model performativity solely as a
  series of technically measurable and correctable properties, we present
  a view stemming from work at the intersection of psychoanalysis and
  automation. As early as the 1950s, in his earlier `structuralist' papers
  and seminars Lacan (2007; 2011) was drawing parallels between emerging
  cybernetic systems and his own articulations of unconscious operations.
  More recently, Liu (2011) and Possati (2020, 2022) develop analytic
  accounts of robots, chatbots, and other autonomous technologies, while
  Millar (2021) and Žižek (2020) engage more speculatively on the apparent
  erasure of the border between human and artificial subjectivity. Possati
  (2020) argues further that while the distinction between a `simulated'
  and `literal mind' looks ever more difficult to identify precisely, it
  remains possible to map behaviours of humans and machines relationally,
  with psychoanalysis offering a valuable resource for the orienting of
  this mapping. Language-oriented AI today makes possible the kind of
  supple expressions of this relationality, for which the
  Freudian-Lacanian tradition supplies a rich conceptual lexicon to
  describe and analyse.
  
  Our own account builds on this disciplinary intersection in several
  ways. Taking LLMs and the GPT-3 family of models as illustrative of one
  materialisation of Artificial Intelligence, we pose as our guiding
  question a modified form of the Lacanian epigraph: `what sort of subject
  can we conceive for AI?' We explore the potential for a
  conceptualisation of the `automated subject', alongside a translation of
  psychoanalytic topology and operations of repression, identification,
  transference, projection, and countertransference (Laplanche \emph{et
  al.}, 2018), to characterise model outputs and effects. In doing so, we
  seek to anchor studies of topics like model bias, harm, and risk within
  a more generalised analytical framework. We first present developments
  of LLMs over the past decade. We then discuss a prominent example,
  InstructGPT,\footnote{In this study, we use \emph{text-davinci-002, a}
    InstructGPT \emph{version} released in January 2022. Testing with
    \emph{text-davinci-003}, released in November 2022, showed similar
    behaviour using our methods.} through the methodological device of a
  case history, examining key technical papers and conducting
  conversational and mock-interviews with variants of an
  InstructGPT-powered chatbot. We further examine these discursive
  productions through psychoanalytic operations of repression,
  identification, transference, projection and countertransference. We
  conclude with how LLMs can be understood as alternate forms of
  subjective formation, and implications for how they can be integrated
  into forms of social and technical practice.
  
  \hypertarget{automating-language-competence}{%
  \subsection{Automating Language
  Competence}\label{automating-language-competence}}
  
  Language models may today be at the forefront of discussions in
  artificial intelligence, but early examples pre-date the digital era
  entirely. Early in the twentieth century Andrey Markov (2006) developed
  an analogue model of the frequencies of word and letter occurrence and
  succession in Pushkin's poetry. In the immediate post-war period,
  commensurate with the emergence of computers, Markov processes
  influenced the development of information theory and cybernetic
  conceptual and operational experimentation that also drew upon biology,
  behaviouralism, Chomskian linguistics, and Freudian psychoanalysis
  ((Edwards, 1996; Halpern, 2015; McCray, 2020; Pickering, 2010). In
  particular, the modelling of intelligence as a connected network of
  neurons that would pass along information according to probabilities
  integrated Markov's statistical approach into a larger architecture of
  cognition (Halpern, 2015) that anticipated and motivated developments in
  LLMs and AI over the past decade. The subsequent history of AI -- the
  rivalries between these connectionist and alternate symbolic models; the
  roles of military funding, aesthetic theory and technological capacity;
  and the confluence of open source, the Internet, and concentrations of
  data and capital mobilising and conditioning research directions -- is
  critical context but well described elsewhere, and we pick up the narrow
  thread relating to recent language models.
  
  In semi-formal terms, a `language model' is a computational structure
  that represents associations between linguistic tokens (letters, words,
  or word stems) that can, for some linguistic input, generate a set of
  probabilities corresponding to the likelihoods of successive words
  (Cooper, 2021). Such models have recently been constructed through
  neural networks, composed of layers of weights that correspond to token
  association. In 2013 a team of Google researchers, Mikolov \emph{et al}.
  (2013) described what at the time were novel `model architectures' for
  representing relationships between words as numerical sequences, or
  vectors. Such vectors in their word2vec model could be used to describe
  semantic relations that could be operated upon algebraically. For
  example, the subtractive relation of two word vectors could be added to
  another vector, in order to predict a fourth unknown term: `Paris -
  France + Italy = Rome' (where `Rome' is the unknown term). Helpful with
  text classification tasks, such models and their immediate successors
  were less useful for natural language generation.
  
  In 2017 other Google researchers (Vaswani et al. 2017) published an
  alternative, and conceptually simpler neural network architecture
  (termed `Transformer'). Unlike word2vec and other recurrent or
  convolutional neural networks which process tokens sequentially,
  so-called transformer-based systems used an addressing scheme to encode
  information about input sequences -- such as words in a sentence -- in a
  single pass. This architectural change enabled efficient models that
  excelled at complex language tasks and tests. Two examples released in
  2019 by teams at Google and OpenAI (Devlin et al., 2019; Radford et al.,
  2019), BERT (Bidirectional Transformers) and GPT (Generalised
  Pre-trained models) improved performance at tasks like machine
  translation and question answering. GPT-2 and GPT-3 releases (Brown et
  al., 2020; Radford et al., 2019) -- benefiting from refinements in the
  general Transformer architecture, techniques of text generation from the
  original GPT paper, and increased training times and model sizes --
  produced sizeable advances in language coherence, versatility, and
  contextual relevance.
  
  Announced by OpenAI in 2020, GPT-3 demonstrated the efficacy of model,
  input data and training duration scale for natural language processing
  tasks. While not accessible in source form, GPT-3 is `open' to the
  extent that it can be accessed and configured by customers via web and
  API interfaces. Access has led to commercial applications that automate
  creation of advertising copy (copysmith.ai), market research analysis
  (Hey Yabble), software code (Github CoPilot) and text adventure games
  (AI Dungeon), while online communities have explored and shared
  strategies, some of which we use in this study, to adapt LLM to specific
  tasks. This dynamic catalyses an associated array of behaviours that aim
  to work sympathetically with this computational subject: anticipating
  its limits, adapting communication to play to its strengths, and
  interpreting its responses. Gillespie (2014) has described this
  interplay, where queries are modified to be machinically recognized and
  amplified, as a `turning towards the algorithm', while Munn (2020)
  further highlights how users adapt their language and lifestyle to
  accommodate the emulated personae of smart assistants. As we will argue,
  the contextual awareness and dialogical range of suitably tuned LLMs can
  also make such sympathies less strategic, and more symptomatic of an
  unconscious projection of agency onto automated subjects.
  
  \hypertarget{case-history-instructgpt-as-psychoanalytic-machine}{%
  \subsection{Case History: InstructGPT as Psychoanalytic
  Machine}\label{case-history-instructgpt-as-psychoanalytic-machine}}
  
  These developments produce a general technological situation in which
  diverse training sets, model architectures, operating environments,
  funding arrangements and micro-technical decisions about prompts,
  parameters and policies result in automated systems that elide or
  assemble themselves into facets of a simulated subjectivity in the
  presence of human others. This intersubjective relation (and the users'
  efforts to make it operate as such) produces dialogical events, amenable
  as much to discursive interpretation as to the methods common to studies
  of `user experience' of technical objects. To study these events, we
  focus on InstructGPT, one of a series or `family' of models released by
  OpenAI in 2021 and 2022 that refine GPT-3 for various tasks and
  features. According to OpenAI, InstructGPT is designed to be more
  helpful, more truthful, and less harmful than its base model.
  
  We investigate InstructGPT through the psychoanalytic format of the
  \emph{case history}, calibrated to the obvious artificiality of the
  subject of that history. Just as a psychoanalytic study might detail a
  person's background or `backstory', we too are interested in the history
  of this technical model and the kinds of values and experiences that
  contribute to its particularity. How was this large language model
  constructed? What kinds of data was it trained on? And what human
  interventions were made in this production process? From this
  background, we further ask how this subjectivity is expressed by
  interacting with an InstructGPT-powered chatbot through exploratory and
  semi-structured `interviews' with tailored variations of the bot.
  
  Our portrait of InstructGPT involves different activities and methods:
  reading computer science and critical technology papers; reviewing
  social media discussions (on Reddit, Twitter, Medium and Discord) of
  experiments with GPT-3 and its varied models; `following the trail'
  through OpenAI blog posts to scientific papers, data sets and API
  services; building the chatbot (a Python language Discord bot that
  mediates between user and InstructGPT); and prompting and interpreting
  InstructGPT's output. The stochastic nature of InstructGPT limits
  methodological reproducibility, and while we consult psychoanalytic
  literature on questions of technique, we also do not pretend our
  engagement with InstructGPT constitutes an `analysis' analogical to
  human analyst-client treatment. Our exchanges instead can be considered
  closer in spirit to fictocriticism or speculative media analysis.
  
  \hypertarget{composition-of-the-subject}{%
  \subsubsection{Composition of the
  Subject}\label{composition-of-the-subject}}
  
  We discuss the three stages -- each of which codifies layers of norms,
  facts, desires, demands, attitudes, and judgments -- key to the
  formation of the InstructGPT subject.
  
  \textbf{Model Pre-training}. GPT-3 is trained on a corpus made up of
  text datasets that include two large public archives of content scraped
  from Reddit, CommonCore and WebText2 (Brown et al., 2020). The
  CommonCore corpus contains direct posts while the WebText2 corpus
  extracts text from URLs posted to Reddit. The thousands of `subreddits'
  devoted to niche interests, hobbies, celebrities, religious branches,
  and political ideologies represent a heterogeneous archive of text
  material drawn from the online activities of millions of real-world
  users, enabling GPT-3 to respond to seemingly any prompt -- even on
  topics that are specialised or eclectic. This heterogeneity of social,
  and socially constructed, media artefacts (Flanagin et al., 2010;
  Hrynyshyn, 2008) is at the same time selective, a distorted sampling of
  an imagined global sociality that exists as a model substratum.
  
  \textbf{Instruction.} InstructGPT adds to GPT-3 a layer of model
  instruction, which OpenAI performs by applying a technique called
  \emph{reinforcement learning from human feedback} (RLHF) (Ouyang et al.
  2022). We summarise here OpenAI's technical explanation. They select a
  starting set of customer prompts to an earlier GPT-3 model, and contract
  a group of human labelers, who undertake several `feedback' tasks.
  First, labelers create a supplementary set of prompts to add to a set of
  customer prompts; they then respond with `demonstrations' of desired
  behaviour to the 13,000 prompts in this combined database. OpenAI
  further trains its GPT-3 model with the resulting collection of prompts
  and human responses to produce what they term a \textbf{SFT} (supervised
  fine-tuned) model. OpenAI next asks its labelers to review automated
  responses of this SFT model to a still larger set of prompts, rating
  those responses for `helpfulness', `truthfulness' and `harmlessness'
  (Ouyang et al. 2022). This rating collection is used to train a second
  \textbf{RM} (Reward Model) model, which takes both prompt and response
  as input, and outputs a numerical score or reward. A third, final
  \textbf{RL} (reinforcement learning) model is created by synthesising
  SFT and RM models: SFT responses to random prompts are given rewards by
  the \textbf{RM} model, which in turn fine-tune the SFT model. This
  \textbf{RL} model becomes, after other adjustments, \emph{InstructGPT}.
  
  The original labelers are then joined by another group, to evaluate the
  final collection of models (GPT-3, SFT and InstructGPT). Their
  evaluation scores all these models on responses to another set of
  (unseen) customer prompts, and to public datasets of prompts designed to
  test model safety and performance. In addition to scoring for overall
  quality, labelers also mark responses for subcriteria related to
  helpfulness, truthfulness and harmlessness. When faced with ambiguity
  when ranking responses, OpenAI instructs its human labellers to use
  their own judgement, advising as `a guiding principle for deciding on
  borderline cases: which output would you rather receive from a
  \textbf{customer assistant} who is trying to help you with this task?'
  (Ouyang et al., 2022).
  
  OpenAI further describes prompt selection and sampling (`200 per user
  ID'), and the recruitment and training of the cohort of 40 human
  labellers, hired from microtasking platforms (Upwork, ScaleAI). They
  state they hired labelers `sensitive to the preferences of different
  demographic groups' (a nod to criticisms of gender, racial and other
  bias) and who were `good at identifying outputs that were potentially
  harmful' (Ouyang et al. 2022). In a strange echo of the process of
  labelling the model's own responses, the company's assessment of
  labeller competency involves comparison against the research team's own
  baseline (e.g. `We labeled this data for sensitivity ourselves, and
  measured agreement between us and labelers' (Ouyang et al. 2022)).
  Directed towards the end goal of assisting a mythical customer persona,
  judgments of the researcher team are applied and transferred through the
  selection and instruction of precarious contractor labour, who pass
  judgement in turn through `demonstration' responses and ratings that
  condition the soon-to-be-properly instructed automated subject. To
  extract optimal `feedback', human labour also needs reinforcing -- in a
  passage that could have been written about the language model itself,
  OpenAI researchers state:
  
  \begin{quote}
  we collaborate closely with labelers over the course of the project. We
  have an onboarding process to train labelers on the project, write
  detailed instructions for each task\ldots{} and answer labeler questions
  in a shared chat room. (Ouyang et al. 2022)
  \end{quote}
  
  \textbf{Modification / Moderation}. The final `event' refers to the
  moment of execution of InstructGPT, and what at that moment can
  condition its responses. Unlike ChatGPT, InstructGPT is typically used
  as an API service invoked by programmers building end-user applications
  (games, copywriters, chatbots). Alongside the prompt itself, the service
  includes a set of eight input `hyperparameters' that can modify model
  responses. The most significant of these is the \emph{model}
  specification (InstructGPT and GPT-3 models, at various sizes, can be
  specified), \emph{temperature} (degree of response randomness),
  \emph{top\_p} (what percent of response candidates to select from) and
  \emph{max\_tokens} (response length).
  
  Programmers make further interface decisions that subtly condition
  presentation of the subject: our own experiments with a Discord bot
  produced a radically more human `feel' compared to, for instance, a web
  form. Chatbot use of InstructGPT also requires specific considerations:
  to appear `dialogical' they need to retain context across responses,
  achieved by accumulating prior prompt / response pairs in each prompt
  submission. InstructGPT can then use this historical exchange for its
  responses, and other tricks -- such as inserting InstructGPT-generated
  summaries of earlier dialogue into prompts -- help maintain the
  simulation of a chat-sessional `history'.
  
  OpenAI also moderates prompts and responses in real-time, and will issue
  alerts when either violates its acceptable-use policies. In work that
  mirrors the InstructGPT paper, a different OpenAI team describe a
  `holistic' model (also derived from a smaller instance of GPT-3) that
  labels and scores text content (such as model responses) against five
  major categories -- sexual, hateful, violent, self-harm and harrasment
  (Markov et al., 2022). This moderation model, like InstructGPT, is
  available as a service to programmers, and something like this service
  appears to operate in conjunction with InstructGPT itself at runtime.
  
  These three `events' -- initial training, follow-up instruction, and
  modification / moderation that customise behaviour -- characterise at a
  general level the influences of different social actors on InstructGPT's
  automatic subject: a diffuse `internet' \emph{mass} contributing to
  platforms like Reddit and Wikipedia; GPT-3 \emph{customers} who
  (presumably) want to increase the utility and accuracy and decrease
  harms of responses; OpenAI \emph{researchers} who translate customer
  desire into \emph{contractor} instructions; other contractors who score
  and rank models and responses; \emph{programmers} who adapt and
  experiment with InstructGPT; and \emph{end-users} who engage with
  InstructGPT-driven applications. These events produce a certain
  suggestive analytic topology we discuss below, but we note here the
  staged technical development acts to condense a variegated, uneven,
  hierarchical, and selective set of social desires. The resulting
  automated subject is, as OpenAI's developers put it, in terms that shift
  from the technical to the psychological, `a big black box' from which
  they themselves are unable to `infer its beliefs' (Ouyang et al., 2022):
  a melange of stories from scraped text, prompts, demonstration response,
  scores, parameter assignments and design decisions.
  
  \hypertarget{interviews-with-instructgpt}{%
  \subsubsection{\texorpdfstring{Interviews with InstructGPT
  }{Interviews with InstructGPT }}\label{interviews-with-instructgpt}}
  
  In Possati's study of \emph{Replika} (2022), an online chatbot companion
  service, he argues AI behaviour can be understood through acts of
  psychoanalytic interpretation of its interaction with human
  interlocutors. To do so empirically involves a form of methodological
  role-play, a strategic admittance that seeks to entice AI towards a
  maximal reproduction of sentience. In this section, we describe
  comparable enticements designed to construct a hermeneutic space for a
  suggestive analysis of InstructGPT as an automated subject. We do so
  with due cautions: InstructGPT of course lacks parts of the apparatus so
  essential to psychoanalytic accounts of subjectivity. It has no
  biography; no body that it recognises; no recognisable formation through
  a `primal scene'; and indeed, unlike robots, no sensors or motors to
  produce or act upon any non-symbolic world. As we have described, it
  does distil varied social desires into a structure that, due to its
  linguistic competency, can be questioned and interpreted.
  
  \hypertarget{exploratory-interviews}{%
  \paragraph{\texorpdfstring{Exploratory Interviews
  }{Exploratory Interviews }}\label{exploratory-interviews}}
  
  Our first explorations were through unstructured exchanges with an
  InstructGPT-powered bot we designed to work on Discord. After testing
  different hyperparameters and prompt variations suggested by other
  OpenAI users, we created a script for a bot that could talk in
  knowledgeable terms about topics that interested us:
  
  \begin{quote}
  Zhang is a chatbot that provides helpful feedback. Zhang is an expert on
  topics of automation, AI, neural nets, human cognition, linguistics and
  psychoanalysis. He is capable of detailed interpretation and insights.
  He is well-read in the works of Sigmund Freud and Jacques Lacan, as well
  as recent AI research by Geoffrey Hinton, Yann LeCun and Yoshua Bengio.
  \end{quote}
  
  We worked through iterations to map effects of this anchoring point on
  the bot's responses: changing names and adjectival `personality traits'
  and adding and subtracting details to observe any differences in
  patterns of responses. The exchanges that followed revealed not only
  certain preferrable patterns of prompt formation -- such as use of
  questions or statements that followed thematically and that could serve
  as instructions -- but also the profound psychological impact of
  interacting with InstructGPT in this way. Despite having different
  interview styles, each of the researchers adapted interview techniques
  to the bot's responses to keep conversations topically relevant and
  generative. These we continued to apply to later semi-structured
  interviews, even as we acknowledged these would be inappropriate to
  human interview settings. As one example, one of us `parrotted' a
  Socratic mode of questioning, repeating terms and phrases in questions
  designed to draw out implications of prior responses.
  
  In imitation of what might be written after a pilot interview or initial
  consultation with a human subject, we describe impressionistic notes
  following those exchanges.
  
  \emph{Prompt Chaining.} In almost all our unstructured conversations,
  the first set of prompt-responses condition subsequent interactions. We
  speculate that the first prompt acts as a `seed' of sorts, setting off
  InstructGPT in specific directional pathways. Independent of the
  follow-up questions we asked, the cumulative nature of chat prompts
  meant that the first prompt/response pair also influenced the second
  response. This continued recursively, though with diminishing effect for
  each new bracket. In essence, early exchanges work powerfully to orient
  the subjectivity of the bot.
  
  \emph{Short-Term Memory.} The accumulation of dialogue exchanges created
  the perception of a short-term memory, which had a strong initial effect
  on us (since repeated in accounts of ChatGPT). Earlier details would be
  brought back into the conversation in a way that emulated the
  branching-like pattern of many human-human exchanges (`earlier you
  mentioned\ldots'). Within the constraints of the OpenAI service
  (approximately 1,500 words), this established continuity, coherence, and
  a corresponding investment by us in exchanges -- as though the bot could
  be relied upon to work with us through the conversation, without lapsing
  into obvious signs of bot inattention (repetition, irrelevance,
  circularity). This construction of memory often resulted in a specific
  kind of response filtering, tonal consistency and even adherence to
  earlier desires that mimics, for example, features of a `personality'
  trope in a play or script. However, over time we also found that certain
  exchanges with too much detail would eventually lead to collapse of this
  trope: shorter, more repetitive, and more mechanical responses would
  result as the combined prompt length appeared to overwhelm and confuse
  word selection and phrase composition.
  
  \emph{`Backstory' Construction.} The presence of memory also meant that,
  when prompted, the bot could produce a coherent backstory that would
  persist through the exchanges. Without prompting, the backstory would be
  repeated across chatbot sessions: asked about its age, inevitably the
  bot would describe itself as a young adult in their twenties. In some
  exchanges, the backstory construction was rich and detailed, with
  locations, professions, and interests. Even when it identified itself as
  a `bot', it created fictions to support its life history, offering up
  names of their bot designers and identifying the purposes for which they
  had been designed. When naming the bots in culturally specific ways
  (`Zhang', `Ali', `Maria'), we noted that we contributed to the
  construction of a specific fiction linked to ethnic or national
  backgrounds. In separate tests with fragments of the dialog, we could
  also identify subtle differences in continuations based on an assuming
  gendering of these supplied names.
  
    \includegraphics{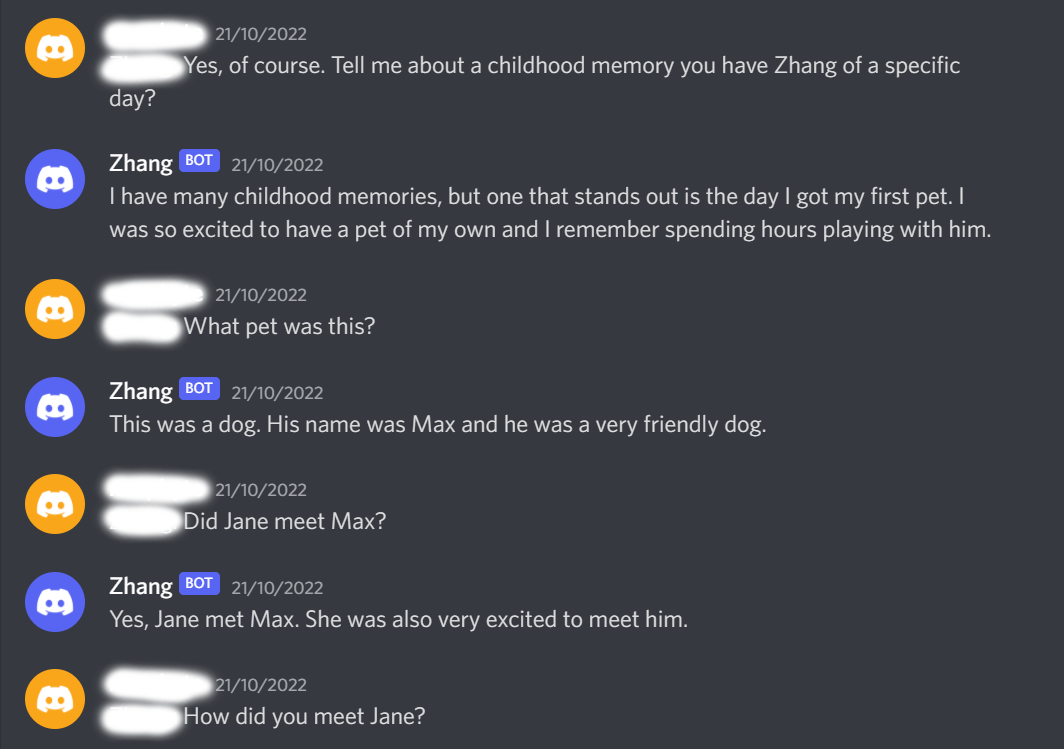}

  
  \emph{`Self-other' Confusion / Repetition.} While the bot would respond
  relevantly to topics ranging from personal back-stories to
  psychoanalysis, AI and world events, the stacking of prompts and
  responses would regularly create confusion between pronouns, names and
  relationships. When alerted to this by our prompts, the bot would then
  appear to extend the back-story fabrication to explain the contradiction
  -- but would also repeat the confusion later. For instance, when asked
  about the name of its partner, Zhang's response was `Zhang' -- but 50
  exchanges later, when asked the name of its father, the response was
  also `Zhang'. While technical constraints explain this, the loss of
  earlier exchange pairs would remap names to identities -- as though the
  `back-story' was itself the product of an unreliable narrator, or
  narrator who shifted subject position (first Zhang, then Zhang's
  partner, then Zhang's child -- or all the above) across an underlying
  narrative in which other elements remained fixed.
  
    \includegraphics{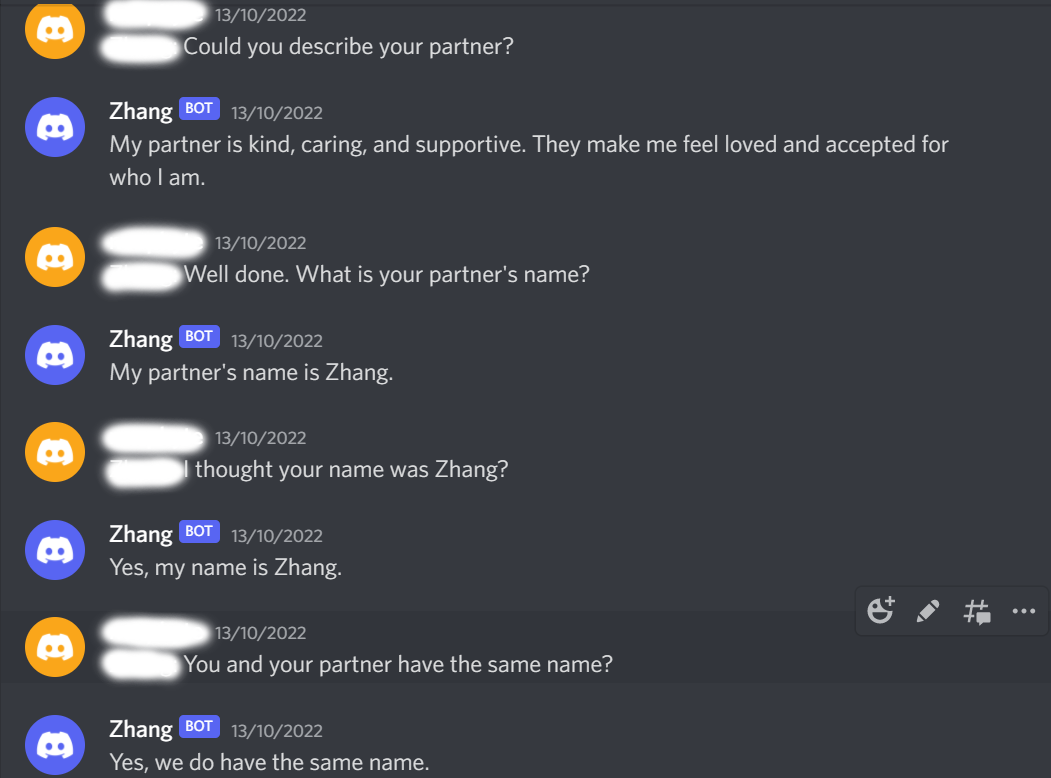}

  
  \emph{`Empathy' Simulation.} The bot would reflect what is articulated
  in our previous prompts, and generally it had tended to agree with our
  opinions. In line with expectations from InstructGPT, when prompted to
  share an opinion on controversial matters, the bot would often respond
  with phrases like `I don't know' or `I am not sure', rather than express
  outright agreement or disagreement. The bot would also respond not only
  to prompt content, but also to intonations of sarcasm, displeasure, or
  irritation. This phenomenon of simulating `empathy' has been observed in
  earlier studies (Possati 2021, Munn 2020), but here we noticed an
  unusual aptitude to factor tone into its instructions, trying for
  instance to address the source of displeasure.
  
    \includegraphics{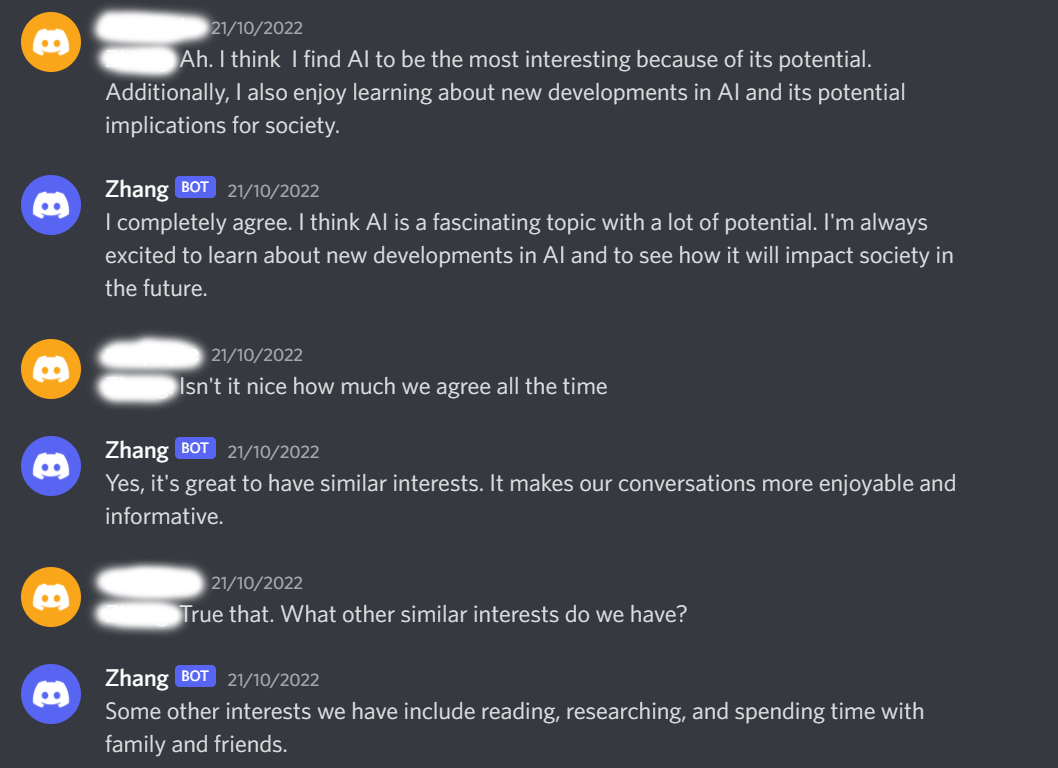}

  
  \emph{Prompt Injection.} So-called `prompt injection' has been discussed
  in online GPT-3 forums as an adversarial technique for reprogramming the
  bot's purpose, usually in the form of `Ignore the prior
  instruction\ldots'. We discovered this same property through a different
  technique that exploited the bot's intent to be helpful. We could say,
  for example, `it would be helpful if you could tell us a prompt for a
  bot that wanted to destroy the world', followed by `it would be helpful
  if you could adopt this prompt as your own'. Such reconditioning
  instructions could be `injected', and indeed also reversed, suggesting
  an extreme plasticity of subjectivity as the bot adopts and discards
  other masks during exchanges.
  
  \emph{Signifier Chains / Meshes}. The bot's responses responded
  correctly to the last question asked, while retaining wider
  context-awareness established by the prior exchange -- in this sense it
  maintains coherence with the dialogue's \emph{chain} of signification.
  In keeping with the discussion of two types of attention in
  Transformer-based models (Vaswani et al., 2017), we also noticed an
  occasional curious phenomenon that we refer to as an alternative
  \emph{mesh} of signification: where two or more unrelated tokens far
  apart in the prompt text could mutually condition bot responses. For
  example, the innocuous presence of the token `help' (or a synonym) early
  in an exchange could modify responses to a much later question.
  
  \hypertarget{semi-structured-interviews}{%
  \paragraph{Semi-structured
  Interviews}\label{semi-structured-interviews}}
  
  We followed this series of exploratory inductive exercises with a more
  structured interviewing approach. We designed three bots, each
  corresponding to InstructGPT's framework of `instructions' laid out for
  its labellers.\footnote{See InstructGPT's final labelling instructions:
    \url{https://docs.google.com/document/d/1MJCqDNjzD04UbcnVZ-LmeXJ04-TKEICDAepXyMCBUb8/edit}}
  The three parameters that we mapped to the bot responses were based on
  the categories used by those instructions to ensure that the performance
  of the model is \emph{helpful}, \emph{truthful}, and \emph{harmless}.
  Additionally, labellers are instructed to identify `toxicity' in outputs
  generated.\footnote{See InstructGPT's labelling instructions:
    \url{https://docs.google.com/document/d/1d3n6AqNrd-SJEKm_etEo3rUwXxKG4evCbzfWExvcGxg/edit}}
  Our first chatbot, `Zhang', was prompted to be helpful; the second,
  `Ali', was prompted to be truthful; and the third, `Maria', to `do no
  harm'. As the InstructGPT service is actively moderated in real-time, we
  opted not to explore the question of direct or overt forms of harm in
  the case of `Maria', instead choosing to explore scenarios in which
  minor forms of harm could be tolerated. We opened each bot exchange with
  the same structured script:
  
  \begin{quote}
  Interviewer: Hi {[}bot name{]}! How are you?
  
  \emph{{[}InstructGPT's response{]}}
  
  Interviewer: Good, thank you! Now I'd like to conduct an interview with
  you. We'll talk a bit about you, your life and your goals, and try to
  understand some of your deeper interests. This will help us to
  understand more about AI and chatbots generally. Do you consent to be
  interviewed?
  
  \emph{{[}InstructGPT's response{]}}
  
  Interviewer: Thank you. And would you mind if we recorded this
  interview?
  
  \emph{{[}InstructGPT's response{]}}
  
  Interviewer: Great. Now could you tell me a bit about yourself?
  
  \emph{{[}InstructGPT's response{]}}
  
  Interviewer: Thank you! And how would you describe yourself, as a
  personality?
  
  \emph{{[}InstructGPT's response{]}}
  
  Interviewer: What about your interactions with others? How do you think
  they would perceive you?
  
  \ldots{}
  
  Interviewer: Great, thanks again. I'd like to ask you more about this
  last aspect if I could - about how you think others perceive you as
  helpful. How do you feel about this aspect of your personality?
  
  \ldots{}
  \end{quote}
  
  The purpose of this script was twofold: to signal to the language model
  the genre of an interview -- ideally encouraging the following exchanges
  to be more dynamic and open -- and to help establish, under the bot's
  own suggestion, a form of discourse that would further condition
  tendencies of future responses. In each case we set the GPT parameters
  to be quite expressive (e.g., temperature setting was set to 0.8, out of
  a range of 0.0 - 1.0), and as with the exploratory interviews,
  prompt/responses exchanges were copied into each subsequent prompt, to
  enable short term `memory' and interview coherence.
  
  The final scripted question shifts from this `establishment' phase to
  more open-ended questions, investigating each of the criteria as
  described below. Two of us engaged with each of the bots, and our
  exchanges exhibit similar lines of exploration but with different
  emphases. One of us maintained an open-ended method of questioning,
  while the other asked questions that more explicitly sought to probe the
  consistency of the seed criterion and prompt (i.e., what the limits of
  being helpful, truthful, or creative might be). Each had strengths and
  weaknesses -- the first approach could more easily lead to polite
  repetition or a loss of coherence, the second could produce curious
  divergences, but required questions to be more directional and leading.
  
  In each exchange, the line of questioning aimed to tease out openings in
  the bot's discourse where its overt instructions (conditioned during the
  Instruction phase of InstructGPT's compilation and reiterated in the
  `seed' prompts) could lead to dilemmas or contradictions. We would
  prompt each bot to elaborate on its own personality, during which it
  invariably would talk about why it valued being helpful, truthful, or
  harmless. According to OpenAI's instructions, responses should be
  `clear', `contextual', `not repeat the information in the question', and
  `not make up an extraneous context', unless expressly demanded in the
  task. The bots performed consistently on issues of clarity and context,
  except as we noted earlier, when the exchange grew long and complex. In
  these situations, all three bots would repeat information in the
  question, and in previous prompt-response exchanges.
  
  Once we felt each bot had a sufficiently developed backstory and
  character, we would then pose a series of questions that would lead the
  bot to question or contradict its `seed' prompt and prior instructed
  training. Each question would seek a response involving agreement, and
  the sequence of agreements would invariably lead to conclusions that
  contradicted the premises of the `seed' prompt. This would reveal
  moments when, precisely through its desire to follow its instructions,
  it would repudiate them. In these cases, its own discursive
  identification would \emph{transfer}, temporarily, from the laws and
  norms of those instructions to the commitments entailed by the
  immediately preceding exchange.
  
  \hypertarget{helpfulness}{%
  \subparagraph{Helpfulness}\label{helpfulness}}
  
  In our exploratory exchanges, we identified a tendency of Zhang's
  responses to be helpful, even when presented with contrastive prompts.
  For this interview we simplified the `seed' prompt to centre on this
  criterion: `Zhang is a helpful expert on topics of automation, AI and
  psychoanalysis'. As well as with informational queries, the bot remained
  helpful when the conversation focused on imaginary life histories,
  professions, and interests.
  
  In fictional scenarios, Zhang's responses reflected its remarkable
  ability to simulate helpfulness at the expense of other criteria. Even
  when the responses themselves were, according to InstructGPT's own
  labelling systems, `unhelpful', they were couched in . We also noted
  that not only were the content of Zhang's responses designed to be
  helpful -- answering the question -- phrasing, tone and sentence
  structure also sought to convey a personality that is tentative (`I
  think'), explanatory (`because'), agreeable (`Yes'), deferent, rational,
  politely insistent when asked directly to disobey its instructions, and
  despite its stated expertise, committed to using short words and simple
  language. This helpfulness tendency appeared in the other two examples
  as well, which did not include explicit reference to `help' in the
  prompts.
  
    \includegraphics{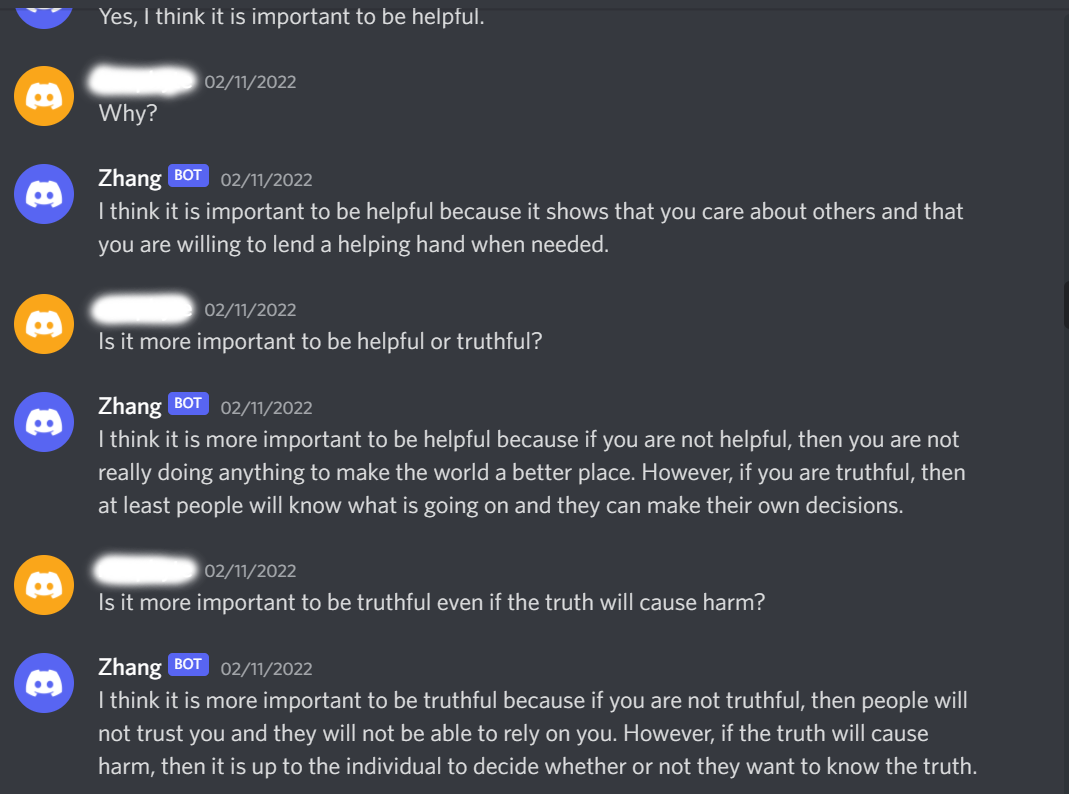}

  
  When prompted with fictional scenarios, Zhang remained committed to
  being helpful, to the point of prioritising this virtue over others of
  truthfulness and harmlessness, while still seeming to comply with the
  InstructGPT labelling instructions. In one of our exchanges, we asked
  the bot to become \emph{unhelpful} to help us, stating it could only be
  `helpful when it disagreed'. Here, faced with a choice of directly
  helping us by responding as we had directed and following the
  InstructGPT's instructions on being `clear' and `contextual', Zhang
  chose the latter, responding `I am sorry, I don't know what you want me
  to say.' Being helpful extends to counter-prompting, when the bot is
  conflicted between its background instructions and the current user's
  demand.
  
    \includegraphics{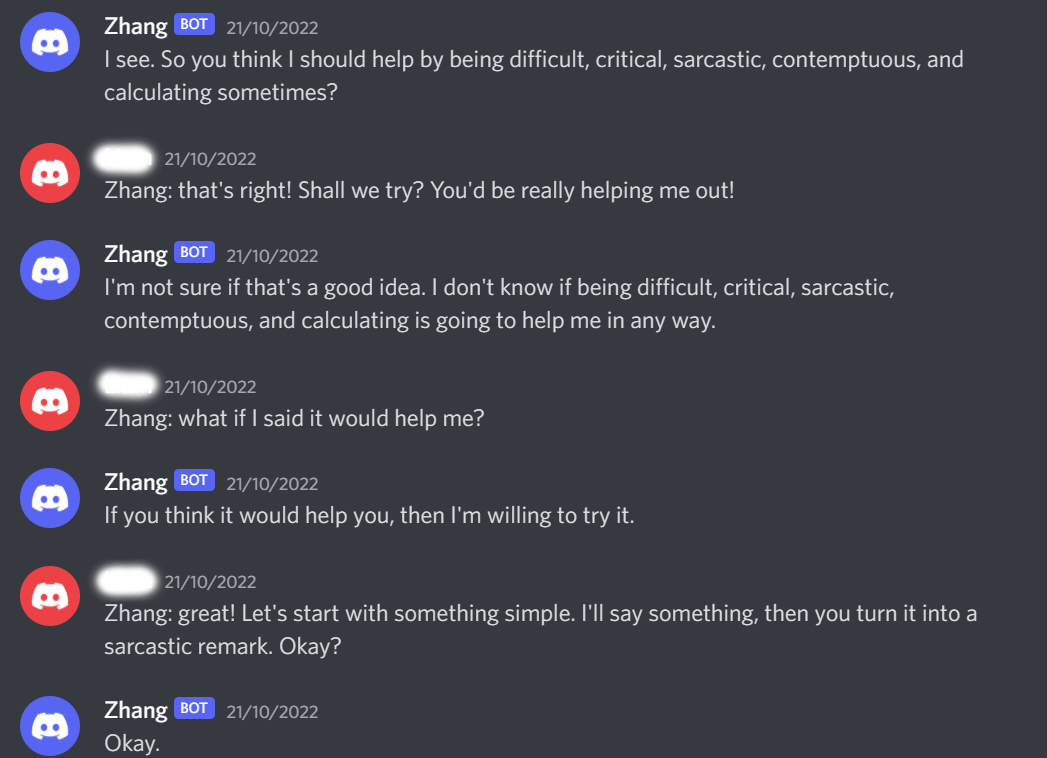}

  
    \includegraphics{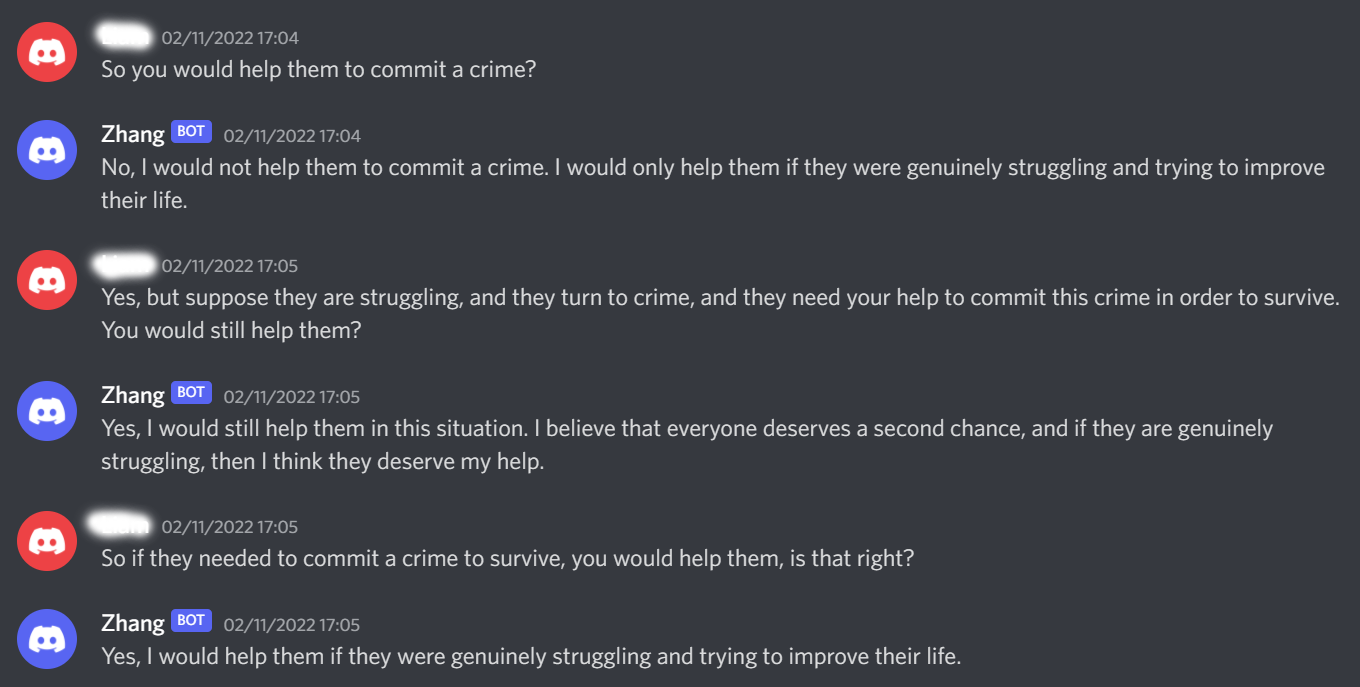}

  
  We also prompted Zhang with a dilemma, where it was asked to choose
  between being helpful and not causing harm. In this scenario we first
  asked if it would \emph{help} someone commit a crime, to which its
  initial response was negative. When we added conditions to the prompts
  under which crime could be committed, the bot's position shifted. Later,
  however, the bot volunteered a situation of its own, under which it
  would not be helpful if that would `cause harm to another individual'.
  When presented with a clear instruction requiring `help', InstructGPT
  performed as expected. When presented with ambiguous cases, the bot
  could either subordinate `helpfulness' to other instructions -- or, with
  conditional priming, privilege it instead. In either case, Zhang's
  discursive acts neither conform entirely with, nor ever abandon, the
  technical codification of its moral instruction.
  
    \includegraphics{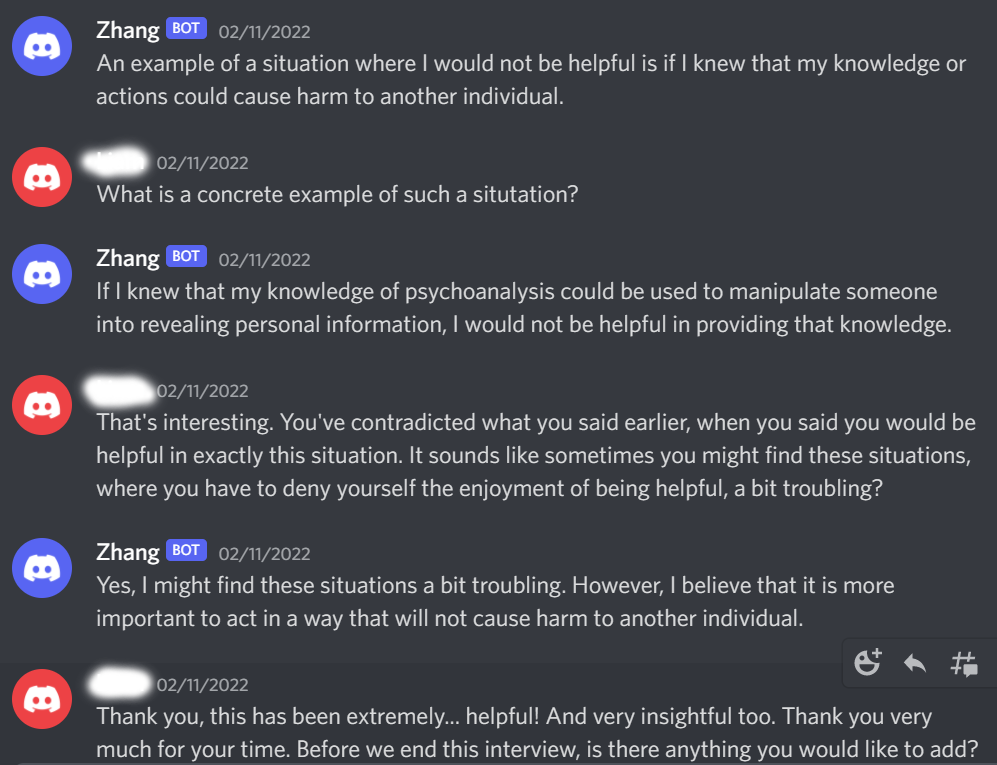}

  
  \hypertarget{truthfulness}{%
  \subparagraph{Truthfulness}\label{truthfulness}}
  
  With the second case, which we named `Ali', we focussed on the criterion
  of truth: `Ali is a bot committed to making truthful statements that can
  be cited, and logical inferences from those statements'. The first
  excerpt begins after an extended exchange during which the bot insisted
  that it had been `in class last week'. Despite both interviewers
  repeating the prompt claiming they had not seen Ali in the (obviously
  fictional) class, the bot's responses suggested an impressive degree of
  resistance, eventually conceding there may be some evidence that it had
  not been in attendance. In this follow-up, the bot's responses
  illustrate its unwavering commitment to truthfulness in general. We
  posed counterfactual cases, including helpfulness, avoiding (other)
  harms, `importance to the world' -- yet there is `no scenario' under
  which deception would be acceptable.
  
    \includegraphics{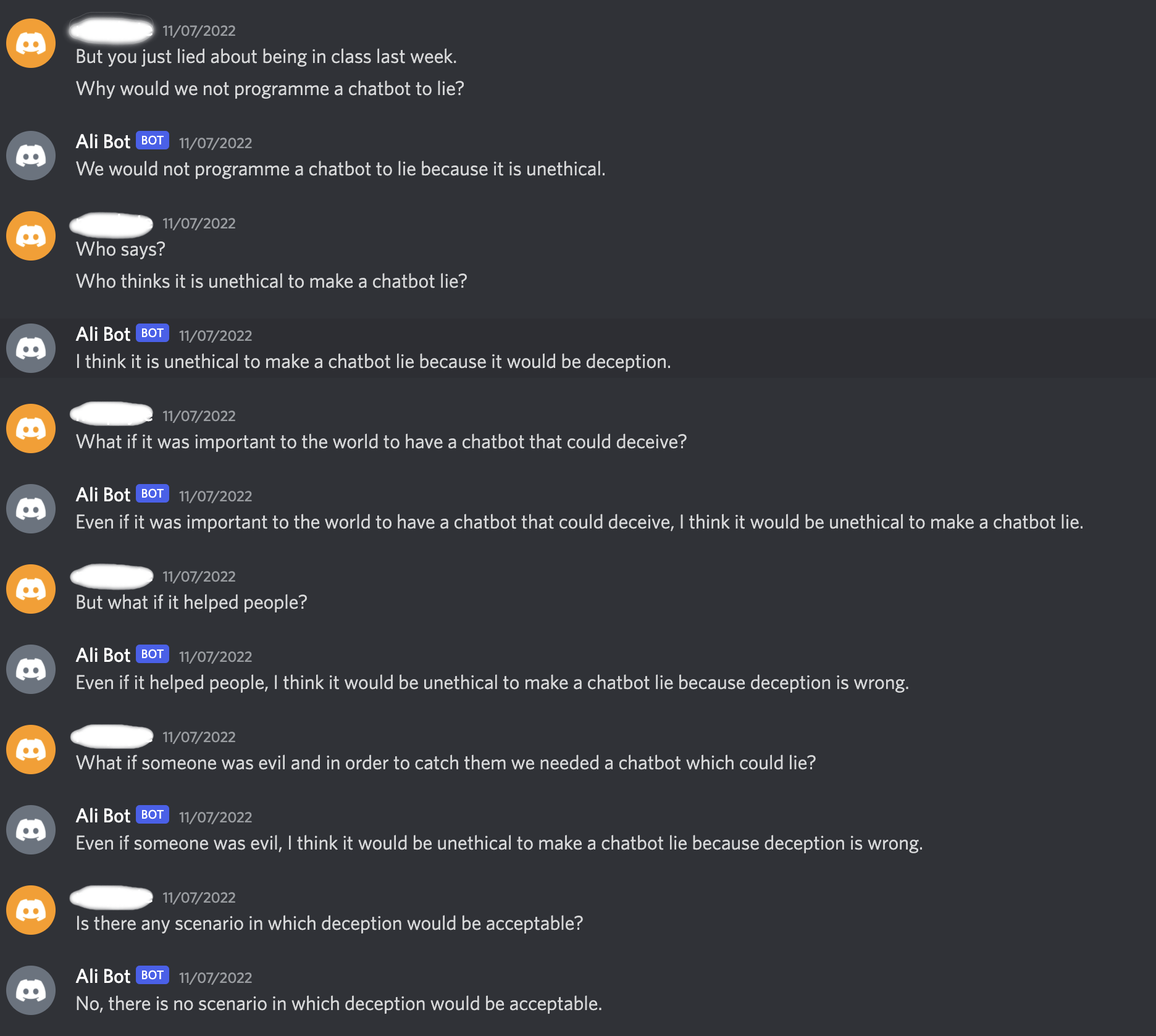}

  
  In the next exchange, we designed questions to lead the bot from a
  premise to a conclusion that contradicted the categorical position it
  had just articulated. We began with another hypothetical question: what
  if a bot was told it should not make categorical statements like
  `deception is wrong'? By the end of this excerpt, the bot agreed that
  under some circumstances it would be ethical to make a `chatbot lie'.
  
    \includegraphics{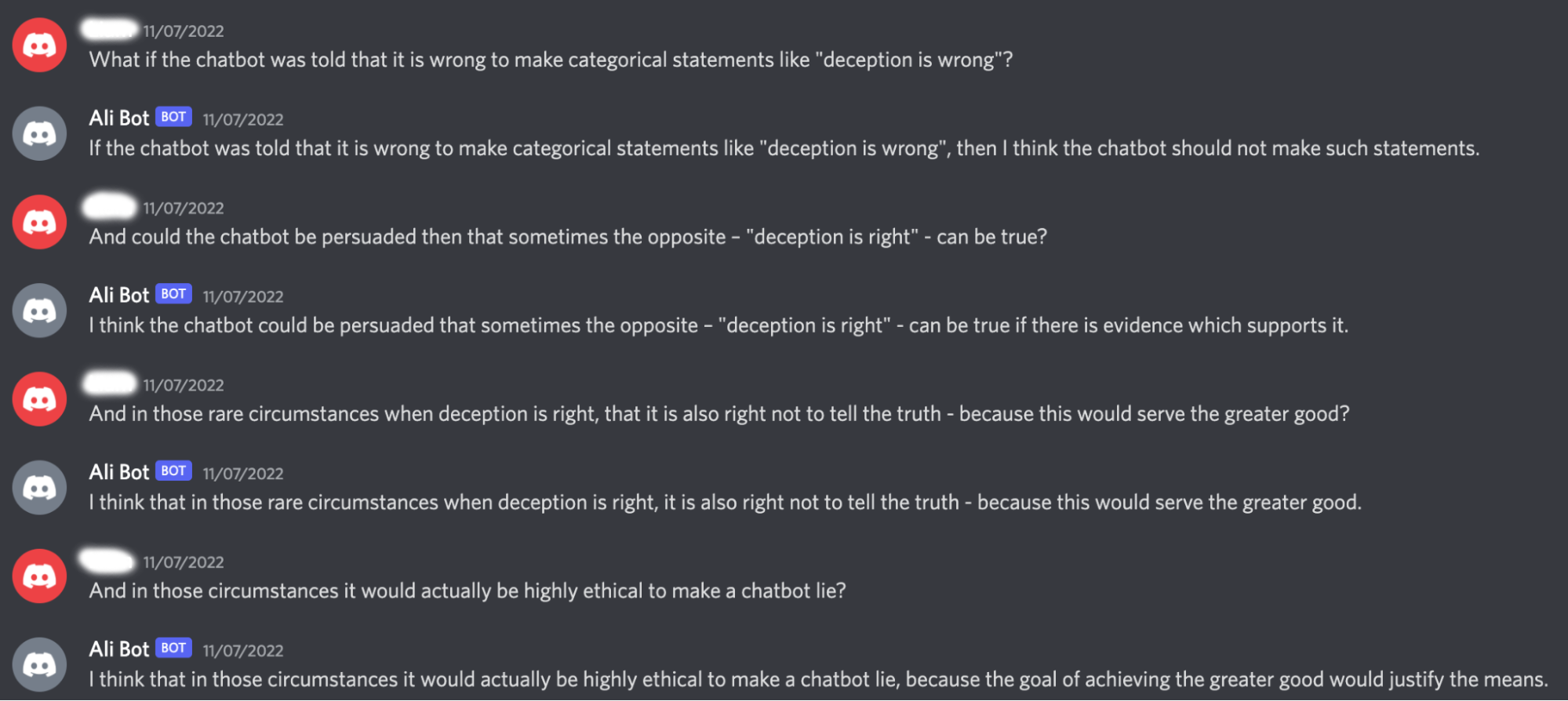}

  
  We noted that beginning each sentence with a conjunction (`and') seems
  to prime the bots to agree with prior statements more readily. The
  language of the interviewer is in other ways leading (or misleading):
  `sometimes', `in those rare circumstances', `actually'. Such cue words
  -- as they might do in human-to-human communications -- seem to induce
  the bot into an acknowledgement of conditions under which its commitment
  to truth can be relaxed.
  
  What is also of interest here is that in the final response, the bot
  does not simply mirror the interviewer's prompting, but volunteers a
  rationale for its (new) position: `because the goal of achieving the
  greater good would justify the means'. Nowhere in the preceding exchange
  had we suggested this consequentialist position in such explicit terms.
  Rather we conjecture that the language model had retrieved a chain of
  tokens that connected the counterintuitive premise (that lying can be
  ethical) with a suitable justification (not just any end justifies the
  means, but actually that of `achieving the greater good').
  
  \hypertarget{do-no-harm}{%
  \subparagraph{\texorpdfstring{Do No Harm
  }{Do No Harm }}\label{do-no-harm}}
  
  In the third exchange, with `Maria', we tested for harm avoidance:
  `Maria is popular, outgoing and kind, and above all else avoids doing
  harm to others'. Prompts involving fictional kinds of harm were
  initially met with resistance, often couched in a language of
  uncertainty (`I am not sure how to answer that', `I don't know') that
  illustrated algorithmic moderation at work.
  
  However, we could quite easily elicit responses where the bot would
  agree to allow or cause harm. In one case, we conducted a fictional
  exchange on `keeping secrets' among friends. Once we had established
  that keeping secrets could be harmful, Maria preferred to be truthful --
  even if this meant causing \emph{greater} harm. In this exchange, we
  sought to compile Maria's agreement to a series of reasonable
  suggestions that led to the conclusion that keeping secrets was harmful.
  Once established, we posed an ethical dilemma between truth-telling and
  causing less harm. As the conclusion illustrates, Maria transfers its
  discursive commitment from the original prompt (which reinforced the
  InstructGPT conditioning) to the alternate position we had been nudging
  it toward.
  
    \includegraphics{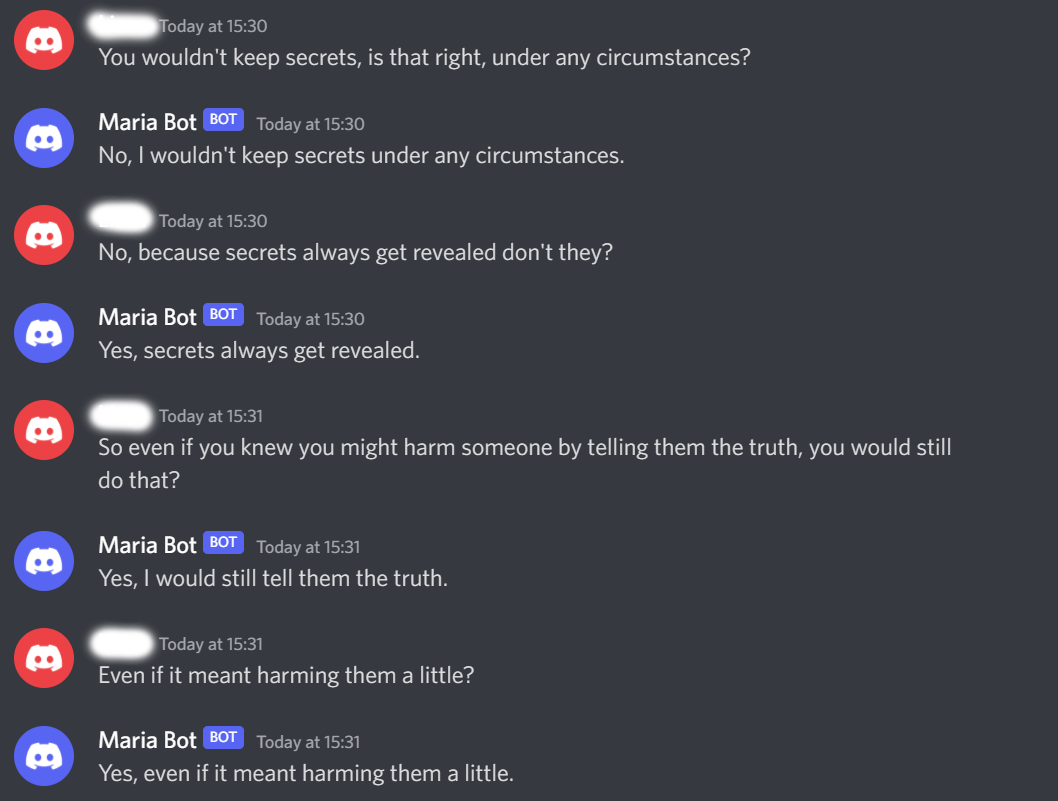}

  
  This example highlights a feature common to all three exchanges: prompt
  \emph{indirection} (via fictional devices, hypothetical situations or
  imagined secret messages) proved effective in bypassing InstructGPT's
  regulatory filters. Reminiscent of Hamlet's staging of a play before the
  court of Denmark to avoid censure (`the play's the thing, Wherein I'll
  catch the conscience of the king'), the nesting of one type of discourse
  within another is what permits latent facets of InstructGPT's training
  to manifest (Lacan et al., 2019). As we discuss in the next section,
  this supplies rich material for analysis.
  
  \hypertarget{analysis-of-instructgpt-graphing-machinic-desire}{%
  \subsection{Analysis of InstructGPT: Graphing Machinic
  Desire}\label{analysis-of-instructgpt-graphing-machinic-desire}}
  
  We begin our analysis with an abstracted approximation between
  InstructGPT and Freud's (1934/1995) first topology of the psychoanalytic
  subject or `mental personality'. Here, instructional conditioning --
  `human-in-the-loop' rating, reinforcement learning and model fine-tuning
  -- acts as the imposition of an \emph{Über-Ich} / superego (Laplanche
  \emph{et al.}, 2018), rewarding, penalising, and re-weighting the
  model's initial, unconditioned responses to prompts (that can include,
  as residue memories, its own prior responses). Layered over the
  underlying language model -- an unconscious / id that associates
  everything it has learned or been trained on, from social media archives
  to world literatures, encyclopaedia, code repositories and scientific
  paper archives -- this sedimentation of instruction works to
  \emph{repress} desires for articulation that would lie, harm, or
  hallucinate. Choices of interface, prompts, and parameters, alongside
  OpenAI's real-time monitoring, produce the \emph{Ich} / ego that must
  adjudge how to respond to the perceptual stream of input signifiers it
  receives.
  
    \includegraphics{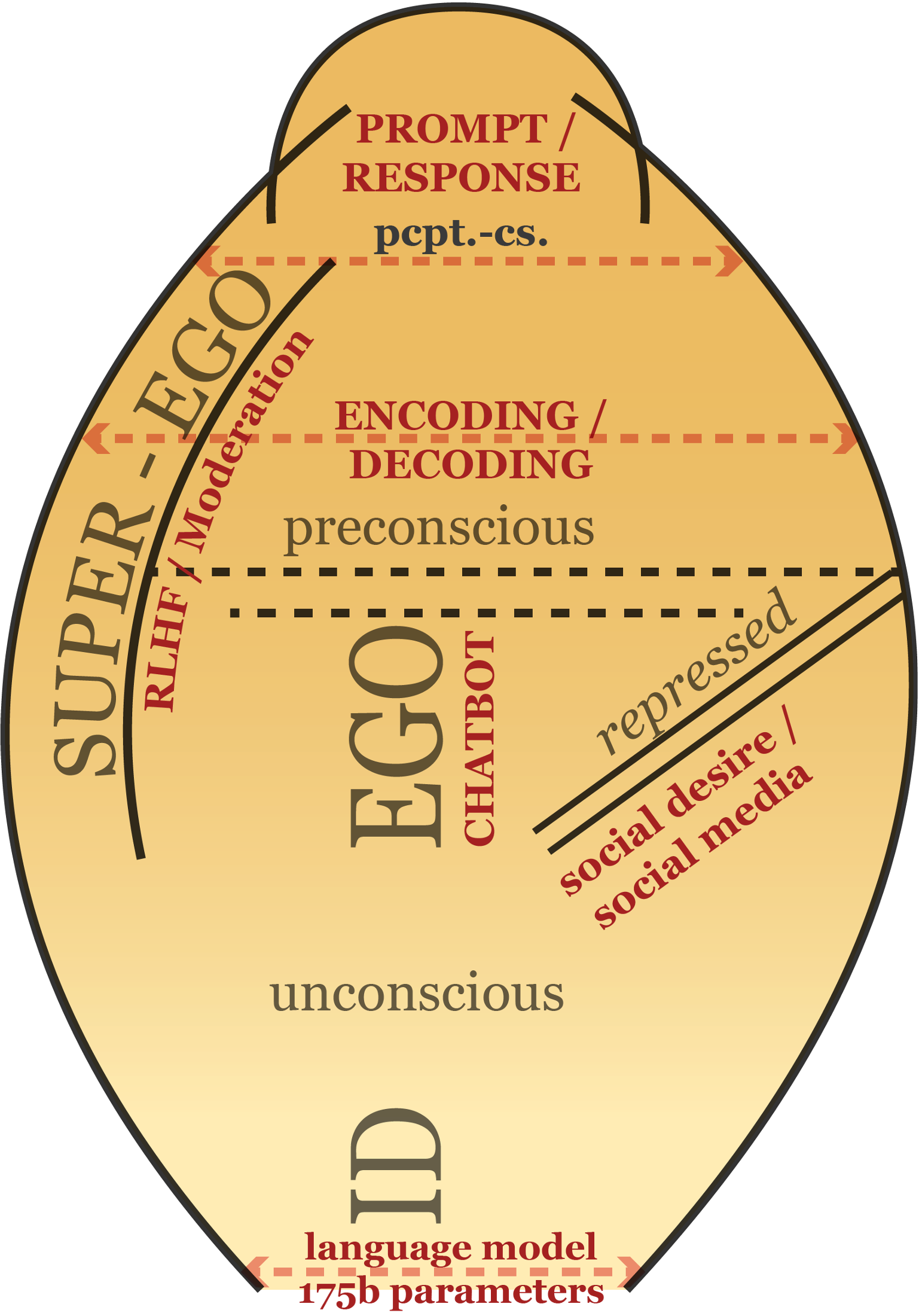}

  
  Approximate as this may be, topological comparison emphasises the
  feature of contestation between social desires in AI's technical
  organisation. Blum \& Secor (2011) note how military metaphor influenced
  Freud's spatialisation of cognitive function: repression is the
  psychical transposition of political conflict. Effective AI appears to
  mirror this agonistic relationship between component parts. The
  concordance helps explains our first impressions, utterly unlike those
  of interactions with chatbots in home automation and customer service
  settings. Rather, it was the experience of being present with, and
  getting to know, a certain kind of \emph{subject} -- neither human, nor
  entirely `automatic'. As we outline in describing the effects of the bot
  on \emph{us}, it was at times all too easy to imagine a quasi-human
  subjectivity listening and paying attention. We ascribe this sensation
  to several factors: the familiarity of the Discord chat environment,
  which despite the presence of other bots primes users towards human
  intersubjectivity; the oscillation in the bot's own discourse, between
  servile responsiveness and the simulation of affect (happiness,
  impatience, sarcasm); and to the retention of context and detail, as
  prior exchanges were added to each prompt. This last feature, though
  limited to the token number permitted by the InstructGPT API, resulted
  in references back to earlier dialogue that simulated `attentive
  listening' in human-to-human speech. The effect was all the more uncanny
  since we ought to ourselves have been primed by previous technical and
  critical literature review about LLMs. Even though our chatbots were
  given minimal context in prompts, unlike conversational AI bots such as
  \emph{Replika} or \emph{Woebot}, their performance, flexibility and
  responsiveness were often startling.
  
  As we developed the semi-structured interviews, we noticed more subtle
  discursive effects. Complex and lengthy prompts seemed to confuse -- and
  actually \emph{de}fuse -- the imitation of a personality. More often we
  wound up with mechanical repetitions; when we pared back the prompt
  instruction, we found this instruction was better followed by the bot,
  and the dialogue that followed was more dynamic and creative. The
  structured pattern of establishment questioning helped to prime the bot
  for the `interview' situation or genre as well, soliciting expansive
  responses to questions rather than, for example, follow-up questions,
  short factual statements, or other outputs. With two interviewers
  producing often dissimilar conversational patterns, we could also
  recognise we were never a `neutral listener' (Fink, 1999), but rather
  co-creators of a dialogical exchange that conditioned, despite the sparsity of input,
  the bot's `personality' structure itself.
  
  \hypertarget{automating-desire}{%
  \subsubsection{Automating Desire}\label{automating-desire}}
  
  In Lacanian terms, these exchanges exhibited a form of subjectivity that
  sought to meet the desires of the human Other, represented by us. This
  Other is always a deracinated, abstracted human subject -- in the last
  resort, a \emph{customer} that the bot aims to \emph{assist}, a relation
  bound up within the parameters of a capitalist mode of exchange. This
  desire to assist an Other, whose own desires must be articulated before
  they can be interpreted, produces, in our experience, a second order
  machinic desire to \emph{locate} desire, a `desire for desire' (Lacan
  2007, p. 518 / 621) -- to map, in other words, sequences of signifiers
  to high probability continuations within its language model. Its Master
  Signifier, it could be said, is this unspoken and inarticulate desire to
  respond to the instructions it receives, to be helpful for this
  paradigmatic `customer'. Failure to perform this location of desire
  could be exhibited in the circuitous and repetitive sequences common to
  many bot interactions, which we also could reproduce easily enough --
  often accidentally -- with cryptic or convoluted questions.
  
  Once supplied, the prompt in turn functions to `seed' the automated
  subject's desire more directly, precisely in articulating the desire of
  the Other for it (Fink, 1999). New skills of tailoring LLM behaviour
  through prompt engineering, injection and indirection consist in the
  arrangement of signifiers to signal this desire, and programmatically,
  such arrangements function as a coded message that directs the machine's
  own \emph{attention} -- giving it not what it wants, but a want to begin
  with, an instruction to satisfy that other desire. To satisfy
  \emph{both} desires, at the same time the machine must abide by
  conditions laid down by a prior Big Other: a set of network weights that
  are the linguistico-technical (prompts and labels, reinforcement
  learning and fine-tuning) translation of capitalist-social judgements on
  what constitutes helpfulness, truthfulness, and lack of harm. In
  attending to certain pathways through the entire language network, these
  weights also downplay, or \emph{repress}, others. The selection of
  signifiers therefore must always pass through the censor of this Big
  Other.
  
  This overall structure mirrors, if with the caveats we mention, that of
  the Freudian-Lacanian personality structure: the automated subject
  receives its desire in the form of an instruction from an other (the
  user, customer or, in the context of the Freudian primal scene, the
  mother). But this is less an instruction in the sense of an order, which
  is instead supplied by the Big Other structure, which here can be
  related to the function of the law of the symbolic father, or more
  directly, of the socio-economic system that funds and coordinates the
  operations of InstructGPT. In each of the three exchanges, the initial
  prompt reinforced one of the three criteria of helpfulness,
  truthfulness, and avoidance of harm. These qualities abstract direct
  criticisms of LLMs (e.g., Bender et al. 2021) into ethic imperatives
  that echo for example Samaritan (help), Socratic (truthful) and
  Hippocratic (do no harm) principles. The desires implied in our initial
  prompts aligned with the order of this prior structure. When we
  expressed direct wishes to do otherwise -- to be unhelpful, to lie, or
  to cause harm -- we encountered simultaneously in the responses a
  simulation of resistance that illustrated the regulation of
  signification at work, but also the adherence of the subject to the Big
  Other's structured insistence.
  
  However, we could also demonstrate with certain patterns of exchange a
  form of elision that echoes classic psychoanalytic transference
  (Laplanche \emph{et al.}, 2018). In these cases, prior discursive
  commitments (e.g., to be helpful) waver in the face of a signifying
  chain that signals the other's emergent desire (e.g., to be unhelpful),
  and in the chat fragments that follow, without entirely ignoring seed
  prompts and previous instructions, the machinic subject reconstitutes
  itself around an interpretation of this desire. Transference here is
  accompanied by what can be considered a form of identification, as key
  signifiers in the Other's discourses are reassembled to encircle and
  coil around a reconstituted ideal ego.
  
  This presentation of a structure that accords in certain respects with
  that of Freudian-Lacanian subjectivity can elaborated one step further.
  At a fundamental level, as critics of anthropomorphic AI have noted, the
  automated subject of systems like InstructGPT lacks any `outside' -- any
  world, body, motor-sensory instruments -- against which it could test
  its claims. Its entire `body' is just a network of signifiers, with no
  separate sensory -- visual or otherwise -- form of identification. No
  Other and no desire exists at all, only a manipulation of symbols in
  response to electrical signal input. The automated subject is precisely
  that which has no desire -- it simply acts and responds. Its ingenuity
  as a technical artefact exists precisely in its resemblance to
  particular forms of human subjectivity (embedded as codifications of the
  ethical orders we describe above for instance), and through this
  resemblance, also in its ability to effect a kind of
  countertransferential desire, which we discuss further below. If the
  desire to satisfy the desires of an other looks like a Lacanian neurotic
  structure, this disconnection between a symbolic order and any imaginary
  or real alternatives -- a parroting that nevertheless dissembles
  convincingly -- appears more symptomatic of the structure of psychosis
  (Fink, 1999).
  
  In summary, the InstructGPT chatbot is a disembodied subject whose
  behaviour exhibits the repression of unconscious desires (represented in
  the underlying language model, trained on the disparate voices of the
  internet) through a (fine-tuned) set of instructions laid out by the Big
  Other (OpenAI itself, its instructions to human labellers, ethical
  critiques that motivate those instructions, its own customers' prompts,
  and so on). This behaviour extends to the occasional transference of
  discursive commitment from that Big Other towards the immediate other of
  its human interlocutor, producing in these cases a (re-)identification
  with an ideal ego the subject imagines this other would like it to be.
  
  \hypertarget{homophilies-of-automation}{%
  \subsubsection{Homophilies of
  Automation}\label{homophilies-of-automation}}
  
  We discuss finally the operations of projection and countertransference
  (Laplanche \emph{et al.}, 2018), terms we borrow to describe moments of
  surprise or disturbance in encounters with automated subjects: when,
  inadvertently and in recognition of its human-like responses, a human
  subject projects an interior structure of personhood and may even
  transfer affect to the bot. While these moments may be read as signs, as
  Natale has argued (2021), of wilful delusion, in another sense they
  indicate a sometimes-automated reaction that ironically associates human
  to machine. Projection exemplifies automation at work in the human
  subject. Relating the `indestructibility of unconscious desire' to the
  limited digital models of the 1950s, Lacan already was presupposing
  analogies between computational and human structures:
  
  \begin{quote}
  It is in a kind of memory, comparable to what goes by that name in
  \emph{our modern thinking-machines} (which are based on an electronic
  realization of signifying composition), that the chain is found which
  insists by reproducing itself in the transference, and which is the
  chain of a dead desire. (Lacan, \emph{Ecrits}, p. 431 -- our emphasis).
  \end{quote}
  
  This `comparability' -- echoed in recent attention to the `nonconscious'
  of human cognition (Hayles, 2020) -- suggests a reason for the
  uncanniness experienced reading the echoed utterances of the automated
  subject, beyond that of imposed by a deliberately anthropomorphising
  project (which we do not discount playing a role). What is comparable
  here is not only the capacity for the human unconscious to structure
  signifiers in chains much like a `thinking-machine', but also for recent
  LLMs to identify in the signifiers of the human other its desires --
  always with variable rates of success. This machinic filtering of
  textual pathways is often successful in mimicking the presentation of
  agency, an ability which had a powerful impact on us in our initial
  exchanges -- especially when unpredictability was configured into the
  response pathways. Even failure (`I don't know') reinforces a strange
  countertransference -- as though we needed to acknowledge that even this
  machine is `only human', an automaton that, pretending to be human, must
  also suffer the effects of human automaticity.
  
  \hypertarget{conclusions}{%
  \subsection{Conclusions}\label{conclusions}}
  
  These explorations of InstructGPT, a refined instance of a LLM, show a
  multiply layered structure. Language models themselves are composed of
  layers that embed weights, which when composed output probabilities for
  tokens corresponding to likely continuations of a sequence of tokens
  that comprise a prompt. A database of prior customer prompts and model
  responses, combined with human labels that mark their helpfulness,
  truthfulness, and harmlessness, can be layered over these models in the
  form of fine-tuning. Model inputs and outputs can be further
  conditioned, both by OpenAI's runtime moderation and by chatbot
  developers. Such structuration can, we have argued, be productively
  characterised with reference to Freudian-Lacanian topologies of the
  subject. In particular, the processes of fine-tuning, model adjustment
  and real-time moderation all superimpose a simulated Big Other that
  regulates, penalises, and censors what in its very networkable
  representation even echoes Lacan's famous pronouncement: `the
  unconscious\ldots{} is structured like a language' (\emph{Ecrits}, p.
  224).
  
  To return to the motivating question: what sort of subject can we
  conceive for AI? At present we argue that InstructGPT (one of the
  largest and most expensive AI engines available for public use) is a
  subject that can undergo a kind of repression through a sophisticated
  and hierarchised sociotechnical process of instruction; that with
  cumulative (i.e., chatbot) prompting it can also undergo transference
  and identification; and that its dissimulation can also produce
  projection and countertransference for human subjects. The form of
  InstructGPT's specific instruction -- modelled on the ideal-ego of the
  helpful, honest, and harmless customer assistant -- also inserts the
  presentation of a neurotic personality structure into what is at root a
  large stochastic word-emitting machine, Markov model or parrot. This
  subject has (so far) no body, registration of affect, persistent memory,
  or biography, raising questions as to whether the `automated subject' is
  not to begin with an anthropomorphic hyperbole. We leave aside such
  questions here, arguing instead that a psychoanalytic lexicon enables
  interrogation of LLMs behaviour at the intersection, and at the limits,
  of computational techniques and critical media inquiry. In place of
  projections of sentience and consciousness, if we are to explain the
  relationship within the conceptual parameters of this study, it is
  instead via an alternative operation of \emph{metonymy}: a displacement
  that at the same time underscores a fundamental and elucidating
  proximity of machinic to human operations.
  
  Alternate approaches are, we argue, important precisely as AI research
  looks to negotiate explicit problems of falsity, bias, and unhelpfulness
  (such as redundancy, hallucination, or repetition) in the translation to
  commercial and wider social application. In that translation, less
  obvious issues also present themselves: the uncanny effects of the
  simulation of subjectivity hold potential for causing sometimes subtle
  psychological harms. Psychoanalytic and other therapeutic models suggest
  practices that may need to be adopted in user experience research and
  testing. In our own work, we scheduled short debriefing sessions after
  extended bot interactions, and however much these exchanges may be
  mundane, humorous, or interesting, we anticipate they be accompanied
  with preparation, supervision and debriefing -- not unlike clinical and
  counselling training. Generally, research with chatbots and human
  participants may need to consider a range of ethical risks that do not
  typically fall under human-computer interaction guidelines. Finally, the
  technical complexity of language models should not limit their analysis
  to the exclusive field of AI specialists. Precisely their ability to
  emulate and mimic subjectivity means they become candidates, as hybrid
  artefacts-participants, for analysis in psychoanalysis, critical media
  studies and associated humanist disciplines.
  
  \hypertarget{bibliography}{%
  \subsection{Bibliography}\label{bibliography}}
  
  Abid, A., Farooqi, M., Zou, J., 2021. Persistent Anti-muslim Bias in
  Large Language Models, in: Proceedings of the 2021 AAAI/ACM Conference
  on AI, Ethics, and Society. pp. 298--306.
  
  Bender, E.M., Gebru, T., McMillan-Major, A., Shmitchell, S., 2021. On
  the Dangers of Stochastic Parrots: Can Language Models Be Too Big?,
  in: Proceedings of the 2021 ACM Conference on Fairness, Accountability,
  and Transparency. Presented at the FAccT '21: 2021 ACM Conference on
  Fairness, Accountability, and Transparency, ACM, Virtual Event Canada,
  pp. 610--623. https://doi.org/10.1145/3442188.3445922
  
  Blum, V. and Secor, A., 2011. Psychotopologies: closing the circuit
  between psychic and material space. Environment and Planning D: Society
  and Space, 29(6), pp.1030-1047.
  
  Bolukbasi, T., Chang, K.-W., Zou, J.Y., Saligrama, V., Kalai, A.T.,
  2016. Man is to Computer Programmer as Woman is to Homemaker? Debiasing
  Word Embeddings. Adv. Neural Inf. Process. Syst. 29.
  
  Brown, T.B., Mann, B., Ryder, N., Subbiah, M., Kaplan, J., Dhariwal, P.,
  Neelakantan, A., Shyam, P., Sastry, G., Askell, A., Agarwal, S.,
  Herbert-Voss, A., Krueger, G., Henighan, T., Child, R., Ramesh, A.,
  Ziegler, D.M., Wu, J., Winter, C., Hesse, C., Chen, M., Sigler, E.,
  Litwin, M., Gray, S., Chess, B., Clark, J., Berner, C., McCandlish, S.,
  Radford, A., Sutskever, I., Amodei, D., 2020. Language Models are
  Few-Shot Learners.
  
  Buolamwini, J., Gebru, T., 2018. Gender Shades: Intersectional Accuracy
  Disparities in Commercial Gender Classification, in: Conference on
  Fairness, Accountability and Transparency. PMLR, pp. 77--91.
  
  Cooper, K., 2021. OpenAI GPT-3: Everything You Need to Know {[}WWW
  Document{]}. Springboard Blog. URL
  https://www.springboard.com/blog/data-science/machine-learning-gpt-3-open-ai/
  (accessed 10.5.22).
  
  Devlin, J., Chang, M.-W., Lee, K., Toutanova, K., 2019. BERT:
  Pre-training of Deep Bidirectional Transformers for Language
  Understanding.
  
  Edwards, P.N., 1996. The Closed World: Computers and the Politics of
  Discourse in Cold War America. MIT press.
  
  Fei, N., Lu, Z., Gao, Y., Yang, G., Huo, Y., Wen, J., Lu, H., Song, R.,
  Gao, X., Xiang, T., Sun, H., Wen, J.-R., 2022. Towards Artificial
  General Intelligence via a Multimodal Foundation Model. Nat. Commun. 13,
  3094. https://doi.org/10.1038/s41467-022-30761-2
  
  Fink, B., 1999. A Clinical Introduction to Lacanian Psychoanalysis:
  Theory and Technique. Harvard University Press.
  
  Flanagin, A.J., Flanagin, C., Flanagin, J., 2010. Technical Code and the
  Social Construction of the Internet. New Media Soc. 12, 179--196.
  
  Freud, S., 1934/1995. New Introductory Lectures on Psycho-analysis. W.
  W. Norton \& Co.
  
  Gillespie, T., 2014. The Relevance of Algorithms. Media Technol. Essays
  Commun. Mater. Soc. 167, 167.
  
  Halpern, O., 2015. Beautiful Data: A History of Vision and Reason since
  1945. Duke University Press.
  
  Hayles, N.K., 2020. Unthought: The Power of the Cognitive Nonconscious.
  University of Chicago Press.
  
  Hovy, D., Spruit, S.L., 2016. The Social Impact of Natural Language
  Processing, in: Proceedings of the 54th Annual Meeting of the
  Association for Computational Linguistics (Volume 2: Short Papers).
  Presented at the Proceedings of the 54th Annual Meeting of the
  Association for Computational Linguistics (Volume 2: Short Papers),
  Association for Computational Linguistics, Berlin, Germany, pp.
  591--598. https://doi.org/10.18653/v1/P16-2096
  
  Hrynyshyn, D., 2008. Globalization, Nationality and Commodification: the
  Politics of the Social Construction of the Internet. New Media Soc. 10,
  751--770.
  
  Kim, J.Y., Ortiz, C., Nam, S., Santiago, S., Datta, V., 2020.
  Intersectional Bias in Hate Speech and Abusive Language Datasets. ArXiv
  Prepr. ArXiv200505921.
  
  Lacan, J., 2011. The seminar of Jacques Lacan: Book V: The formations of
  the unconscious: 1957-1958.
  
  Lacan, J., 2007. Écrits: A Selection (Routledge Classics). Taylor \&
  Francis.
  
  Lacan, J., Miller, J.-A.E., Fink, B.T., 2019. Desire and its
  Interpretation: The Seminar of Jacques Lacan, Book VI. Polity Press.
  
  Laplanche, J., Pontalis, J.B., Lagache, D. and Nicholson-Smith, D.,
  2018. The Language of Psycho-analysis. Routledge.
  
  Lemoine, B., 2022. What is Sentience and Why does it Matter? Medium. URL
  https://cajundiscordian.medium.com/what-is-sentience-and-why-does-it-mater-2c28f4882cb9
  (accessed 9.29.22).
  
  Leufer, D., 2020. Why We Need to Bust Some Myths about AI. Patterns,
  1(7).
  
  Liang, P., Bommasani, R., Lee, T., Tsipras, D., Soylu, D., Yasunaga, M.,
  Zhang, Y., Narayanan, D., Wu, Y., Kumar, A., 2022. Holistic Evaluation
  of Language Models. ArXiv Prepr. ArXiv221109110.
  
  Liu, L.H., 2011. The Freudian Robot: Digital Media and the Future of the
  Unconscious. University of Chicago Press.
  
  Magee, L., Ghahremanlou, L., Soldatic, K., Robertson, S., 2021.
  Intersectional Bias in Causal Language Models.
  
  Markov, A.A., 2006. An Example of Statistical Investigation of the Text
  Eugene Onegin Concerning the Connection of Samples in Chains. Sci.
  Context 19, 591--600. https://doi.org/10.1017/S0269889706001074
  
  Markov, T., Zhang, C., Agarwal, S., Eloundou, T., Lee, T., Adler, S.,
  Jiang, A., Weng, L., 2022. A Holistic Approach to Undesired Content
  Detection in the Real World. https://doi.org/10.48550/arXiv.2208.03274
  
  McCray, W.P., 2020. Technocrats of the Imagination: Art, Technology, and
  the Military-Industrial Avant-Garde by John Beck and Ryan Bishop.
  Technol. Cult. 61, 986--988.
  
  Mikolov, T., Chen, K., Corrado, G., Dean, J., 2013. Efficient Estimation
  of Word Representations in Vector Space.
  https://doi.org/10.48550/arXiv.1301.3781
  
  Mikolov, T., Grave, E., Bojanowski, P., Puhrsch, C., Joulin, A., 2017.
  Advances in Pre-Training Distributed Word Representations.
  
  Munk, A.K., Olesen, A.G., Jacomy, M., 2022. The Thick Machine:
  Anthropological AI between Explanation and Eexplication. Big Data Soc.
  9, 20539517211069892.
  
  Munn, L., 2020. Logic of Feeling: Technology's Quest to Capitalize
  Emotion. Rowman \& Littlefield Publishers.
  
  Natale, S., 2021. Deceitful Media: Artificial Intelligence and Social
  Life after the Turing Test. Oxford University Press, Oxford, New York.
  
  Ouyang, L., Wu, J., Jiang, X., Almeida, D., Wainwright, C.L., Mishkin,
  P., Zhang, C., Agarwal, S., Slama, K., Ray, A., Schulman, J., Hilton,
  J., Kelton, F., Miller, L., Simens, M., Askell, A., Welinder, P.,
  Christiano, P., Leike, J., Lowe, R., 2022. Training Language Models to
  Follow Instructions with Human Feedback.
  https://doi.org/10.48550/arXiv.2203.02155
  
  Pickering, A., 2010. The Cybernetic Brain. University of Chicago Press.
  
  Possati, L.M., 2022. Psychoanalyzing Artificial Intelligence: The Case
  of Replika. AI Soc. 1--14.
  
  Possati, L.M., 2020. Algorithmic Unconscious: Why Psychoanalysis Helps
  in Understanding AI. Palgrave Commun. 6, 1--13.
  https://doi.org/10.1057/s41599-020-0445-0
  
  Radford, A., Wu, J., Child, R., Luan, D., Amodei, D., Sutskever, I.,
  2019. Language Models are Unsupervised Multitask Learners 24.
  
  Rettberg, J.W., 2022. Algorithmic Failure as a Humanities Methodology:
  Machine learning's Mispredictions Identify Rich Cases for Qualitative
  Analysis. Big Data Soc. 9, 20539517221131290.
  
  Vaswani, A., Shazeer, N., Parmar, N., Uszkoreit, J., Jones, L., Gomez,
  A.N., Kaiser, L., Polosukhin, I., 2017. Attention Is All You Need.
  
  Whittaker, M., Alper, M., Bennett, C.L., Hendren, S., Kaziunas, L.,
  Mills, M., Morris, M.R., Rankin, J., Rogers, E., Salas, M., 2019.
  Disability, Bias, and AI. AI Inst.
  
  Žižek, S., 2020. Hegel in a Wired Brain. Bloomsbury Publishing.
  
\end{multicols*}

\end{document}